\documentclass[aps,10pt,twocolumn]{revtex4}
\usepackage[utf8x]{inputenc}
\usepackage{graphicx}
\usepackage{epsfig}

\begin{document}
\title{Anomalous fluctuations of nematic order in 
solutions of semiflexible polymers}

\author{Sergei A. Egorov$^{1,3,4}$, Andrey Milchev$^2$, and Kurt Binder$^3$}

\affiliation{$^1$ Department of Chemistry, University of Virginia,
Charlottesville, VA 22901, USA [E-Mail:sae6z$@$cms.mail.virginia.edu] \\ 
$^2$ Institute for Physical Chemistry, Bulgarian Academia of Sciences, 1113
Sofia, Bulgaria \\
$^3$ Institut f\"ur Physik, Johannes Gutenberg Universit\"at Mainz, 55099 Mainz,
Germany\\ $^4$ Leibniz-Institut f\"ur Polymerforschung, Institut Theorie der
Polymere, Hohe Str. 6, 01069 Dresden, Germany}

\begin{abstract}
The nematic ordering in semiflexible polymers with contour length $L$ exceeding
their persistence length $\ell_p$ is described by a confinement of the polymers
in a cylinder of radius $r_{eff}$ much larger than the radius $r_\rho$, expected
from the respective concentration of the solution. Large scale Molecular
Dynamics simulations combined with Density Functional Theory are used to locate
the Isotropic-Nematic ($I-N$)-transition and to validate this cylindrical
confinement. Anomalous fluctuations, due to chain deflections from neighboring
chains in the nematic phase are proposed. Considering deflections as collective
excitations in the nematically ordered phase of semiflexible polymers elucidates
the origins of shortcomings in the description of the $I-N$ transition by
existing theories. 
\end{abstract}

\maketitle

{\em Introduction}.---The stiffness of semiflexible macromolecules in solutions and melts creates a
tendency towards liquid-crystalline order. However, neither  the precise
conditions for the onset of nematic order, nor the properties of the phases are
well understood.~\cite{ciferri91,donald06,grosberg81,sato96} Semiflexible
polymers behave like rigid rods on the persistence length scale $\ell_p$, yet
like random coils~\cite{degennes79} on larger scales, if their contour length $L
\gg \ell_p$. In solutions of semiflexible polymers good solvent conditions
prevail, the effective monomer-monomer interactions being repulsive. The monomer
concentration $\rho$ is then the control parameter for the onset of order.
Unlike solutions of rod-like particles (e.g., the tobacco mosaic
virus~\cite{fraden89}), where translational and orientational entropy
contributions compete~\cite{onsager49}, here also conformational degrees of
freedom due to chain flexibility matter. This hampers the understanding of such
systems~\cite{khokhlov81,khokhlov82,odijk85,odijk86,chen93,hentschke90,dupre91,
sato90,sato94,fynewever98}: even in the limit $\ell_p \gg d$ ($d$ being the
effective monomer diameter) the extension of Onsager's
theory~\cite{onsager49} for the isotropic (I) - nematic (N) transition of thin
long rods is difficult.~\cite{khokhlov81,khokhlov82,odijk85,odijk86,chen93}
Attempts~\cite{hentschke90,dupre91,sato90,sato94,fynewever98} to go beyond this
limit have produced contradictory results: at the concentrations of interest it
no longer suffices to deal with the inter-chain interactions via the 2-nd virial
coefficient only, 
as~\cite{onsager49,khokhlov81,khokhlov82,odijk85,odijk86,chen93} for 
$\ell_p \gg d$. However, progress in the understanding of these lyotropic crystalline
polymers is highly desirable in view of interesting applications (various liquid
crystal devices~\cite{ciferri91,donald06}, emerging new types of complex soft
materials such as nematic elastomers~\cite{finkelmann01}, nematic
emulsions~\cite{poulin97}, etc.), and in the context of biological matter (the
stiff cytoskeleton networks, neurofilaments within the
axon~\cite{aldoroty87,hirokawa84}, intermediate filaments in
cells~\cite{koster15,huber15}, etc.)

In the present Letter we take steps towards elucidating this important problem
by means of large scale Molecular Dynamics (MD) simulations, analyzing them in
terms of the ``deflection length'' concept. This length, $\lambda$, was
originally used to describe confinement of semiflexible chains in cylindrical
tubes~\cite{odijk83,yang07,chen13,chen15}. We will explain why long wavelength
collective fluctuations occur, causing large deflections of the polymers from
their director. The observed reduction of the nematic order parameter $S$ is
then stronger than predicted by Density Functional Theory (DFT), even when the
DFT prediction for the location of the transition is validated by
MD.~\cite{notedft}

Simulations of $I-N$ transitions have been attempted 
earlier~\cite{wilson93,dijkstra95,escobedo97,vanwesten13,weber99,ivanov03, ivanov07},
albeit only short chains and small simulation boxes could be handled (we
disregard thereby lattice models~\cite{weber99,ivanov03,ivanov07} where the
chains can order only in discrete directions and no deflection length exists). In
the present work both $\ell_p$ and the chain length $N$ (i.e., the number of
beads in chains) are widely varied, $8 \le N \le 128$, and large systems (up to
$500000$ beads) have been used. Our work has become feasible by means of very
efficient codes~\cite{anderson08,glaser15} on graphical processing units 
(GPUs).~\cite{notegpu}

{\em Model}.---We employ the standard model~\cite{kremer90} where beads interact along the
chains with the spring potential $U^{FENE}(r)$ while any pair of beads interact
with the repulsive part of the Lennard-Jones (LJ) potential, $U^{LJ}(r)$, $r$
being the distance between beads. In this model the distance between the neighboring beads along the chain is $\ell_b = 0.970 \sigma$ (hence the
contour length $L=(N-1) \ell_b$), with the LJ diameter $\sigma=d=1$ and the LJ energy 
$\epsilon=1$, as well as temperature $T=1$; the integration time step is $\delta t =
0.01$ ($\tau = \sqrt{m\sigma^2/\epsilon} = 1$ MD time unit)~\cite{notemd}. 
Chain stiffness is described by the bond bending potential 
$U^{bend}(\theta_{ijk}) = \epsilon_b [1 - \cos (\theta_{ijk})]$ for $j=i+1, \;k=j+1$.
Here $\theta_{ijk}$ is the angle between the bond vector $\vec{a}_i = \vec{r}_j
- \vec{r}_i$ and $\vec{a}_j = \vec{r}_k - \vec{r}_j$. The persistence length is
then simply~\cite{hsu10} 
$\ell_p/\ell_b = -1/\ln \langle \cos(\theta_{ijk})\rangle =\epsilon_b$ 
for $\epsilon_b \ge 2$. 

\begin{figure}[htb]
\vspace{0.2cm}
\hspace{-1.0cm}
\includegraphics[scale=0.26]{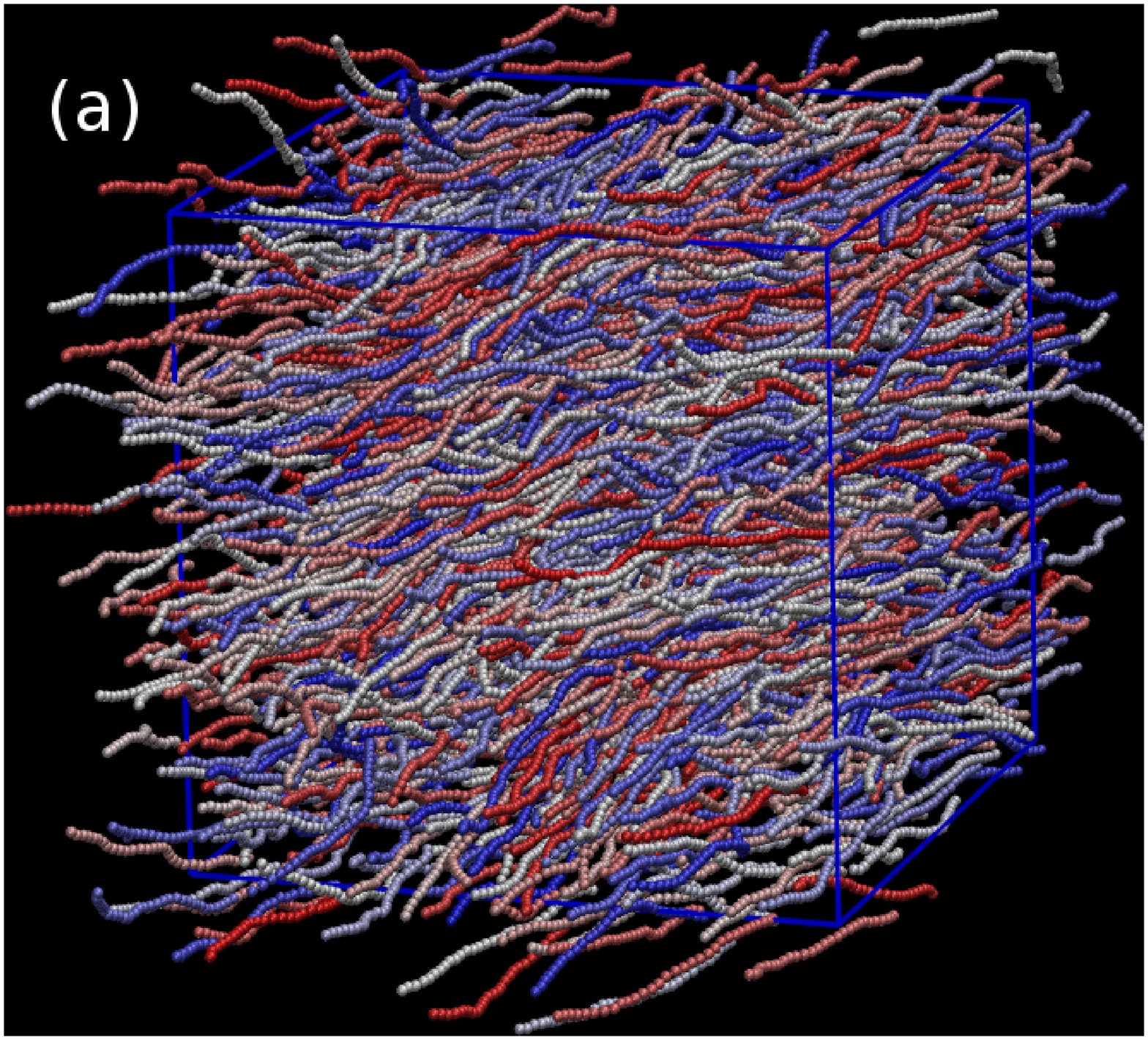}

\vspace*{0.2cm}
\includegraphics[scale=0.19]{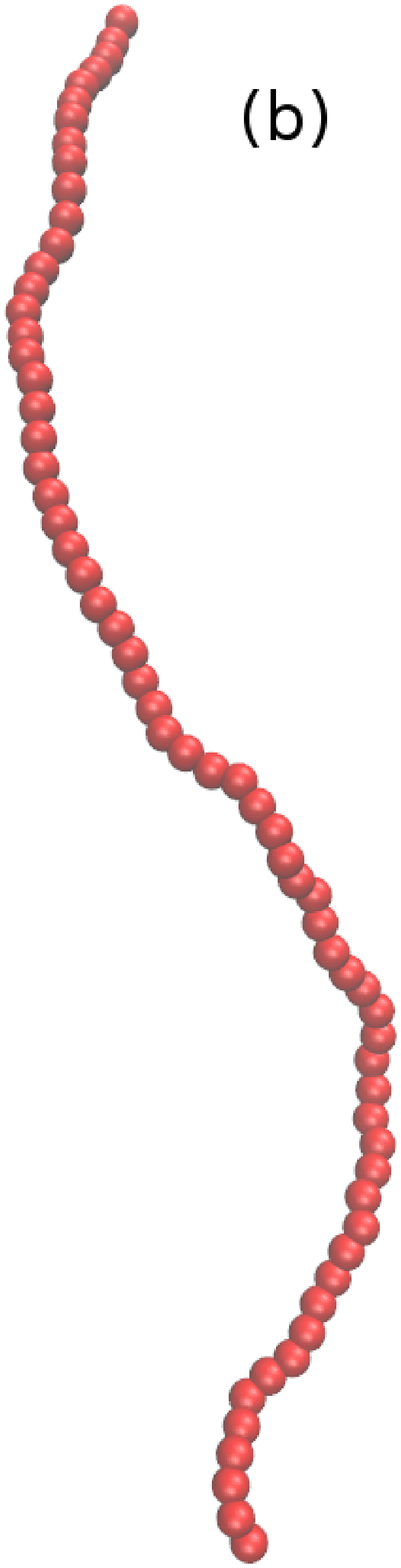}
\hspace{0.7cm}
\includegraphics[scale=0.26]{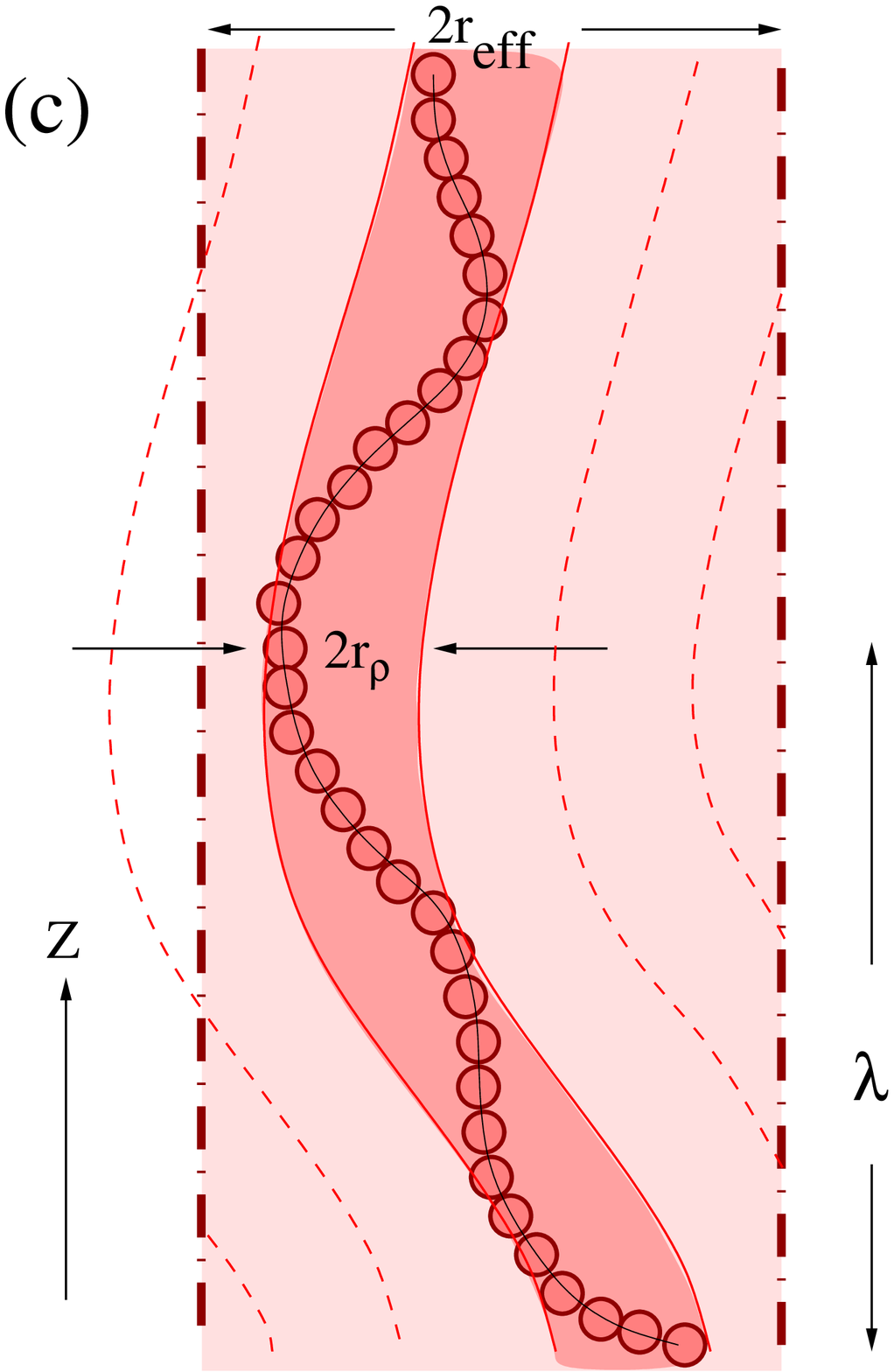}

\caption{(a) Snapshot of a system of semiflexible polymers with length $N=32$,
stiffness $\epsilon_b=100$, at concentration $\rho=0.6$ (deep in the nematic
phase). (b) Typical conformation of a semiflexible polymer in the nematic phase
($N=64,\; \epsilon_b=16,\; \rho=0.4$). (c) Schematic description of nematic
order: each chain has its own cylindrical (bent) tube of diameter $2 r_\rho$,
defined such that it contains only monomers from the considered chain. The tube
is placed inside a straight wider cylinder of diameter $2 r_{eff}$ (see text).
The definition of the deflection length $\lambda$ is indicated.}
\label{fig1}
\end{figure}

We vary $\epsilon_b$ from $\epsilon_b = 8$ to $\epsilon_b = 128$. The linear
dimension $L_{box}$ of the cubic simulation box (with periodic boundary
conditions) is chosen large enough so that even fully extended chains still fit
in. Thus, with ${\cal N}$ chains altogether, the monomer concentration $\rho = {\cal N}N/L_{box}^3$.
System trajectories are computed with the velocity-Verlet algorithm, applying as
usual a Langevin~\cite{kremer90} thermostat. 
 Pressure $P$ is computed using the Virial
theorem, and the order parameter $S$ is the largest eigenvalue of
the tensor $Q_{\alpha \beta}$ which describes the average orientation of the
unit vectors along bonds in the system.~\cite{egorov16} 

Fig.~\ref{fig1}a shows a typical configuration in the nematic phase. Although
there the value of $S$ is large ($S\approx 0.9$), considerable bending of the
wormlike chains is observed. On the molecular scale, the character of this phase
differs considerably from a nematic formed by rod-like molecules. Long
wavelength excitations (deflections of chain orientation around the common
director) are clearly seen from typical configurations of individual chains as
the simulation snapshots prove (Fig.~\ref{fig1}b). These observations suggest a
more comprehensive coarse-grained picture of nematic order in solutions of
semiflexible polymers (Fig.~\ref{fig1}c) that we explain below. 

\begin{figure}[htb]
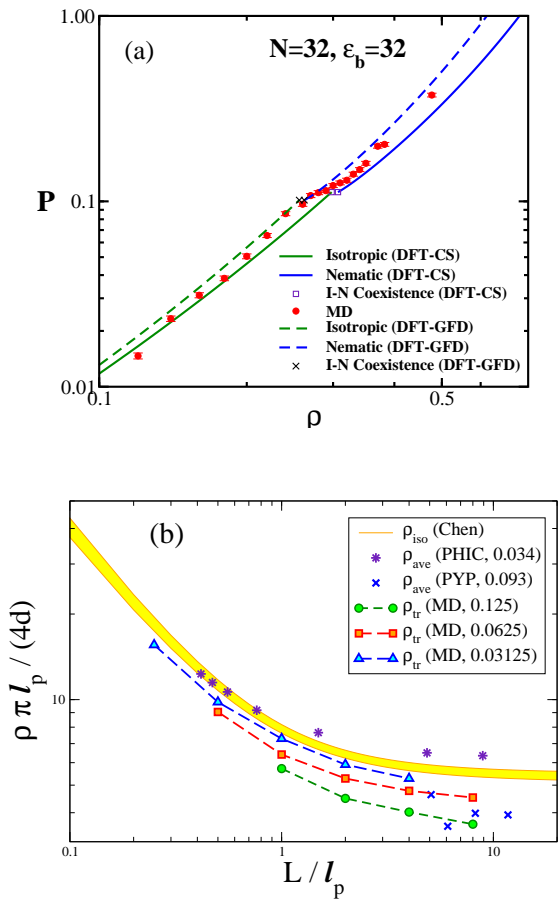

\includegraphics[scale=0.3]{figprsn32eps32.eps}

\vspace*{0.85cm}
\includegraphics[scale=0.3]{figexpt.eps}

\caption{Pressure vs concentration for the case $N=32,\;\epsilon_b=32$,
according to MD (full dots) and two versions of density functional theory,
DFT-CS (solid lines) and DFT-GFD (broken lines)~\cite{notedft}. The $I-N$
transition in the simulation is rounded by finite-size effects. DFT predictions
for $I-N$ coexistence are indicated by squares (DFT-CS) and crosses (DFT-GFD).
(b) Scaled volume fraction $\rho\pi\ell_p/(4d)$ at the transition plotted versus
$L/\ell_p$ according to MD, theory~\cite{chen93} and typical
experiments~\cite{itou88,abe91}. The shaded stripe indicates the I-N coexistence
region, $\rho_i<\rho<\rho_n$, as predicted by Chen~\cite{chen93}. MD does not
resolve $\rho_i,\rho_n$, rather $\rho_{tr}$ is the position of the maximum slope
of the $S$ vs $\rho$ curve (see Fig.~\ref{fig3}). For the experiments, namely
poly(hexyl isocyanate) (PHIC) in toluene~\cite{itou88} and poly(yne)-platinum
(PYP) in trichloroethylene~\cite{abe91} $\rho_{ave}=(\rho_i+\rho_n)/2$ was taken
as the transition density. The numbers in the brackets in the legend indicate
$d/\ell_p$.} \label{fig2}
\end{figure}

{\em Equation of state and order parameter}.---Typical data for pressure
(Fig.~\ref{fig2}a) versus concentration reveal qualitative agreement between MD
and DFT. Of course, the latter cannot use the continuous potentials used by MD,
but rather is based on extensions of a tangent hard-sphere chain
model~\cite{fynewever98,egorov16}, and different choices of the equation of state
within the DFT framework yield slightly different results (see Supplementary
Information). Thus, perfect quantitative agreement between the DFT prediction
for the location of the $I-N$ transition and the simulation cannot be expected.
Interestingly, Fig.~\ref{fig2}b reveals a similar trend as the
Khokhlov-Semenov-Odijk-Chen~\cite{khokhlov81,khokhlov82,odijk85,odijk86,chen93}
theory, when we plot the volume fraction $\rho_{tr}\pi/4$ at the transition
multiplied by $\ell_p/d$ versus $L/\ell_p$. However, unlike usually
assumed~\cite{sato96,khokhlov81,khokhlov82,odijk85,odijk86,chen93,dijkstra95},
this does not yield a universal master curve, but rather a decrease of
$\rho_{tr}$ (at fixed $L/\ell_p$)  with increasing  ratio $d/\ell_p$ takes
place. In fact, this finding helps understand the origin of discrepancies
between
theories~\cite{khokhlov81,khokhlov82,odijk85,odijk86,chen93,hentschke90,dupre91,
sato90,sato94,fynewever98} and experiments~\cite{sato96,abe91,itou88} where a fit of all systems to a universal master curve was assumed~\cite{sato96,dijkstra95}.

\begin{figure}[htb]
\includegraphics[scale=0.3]{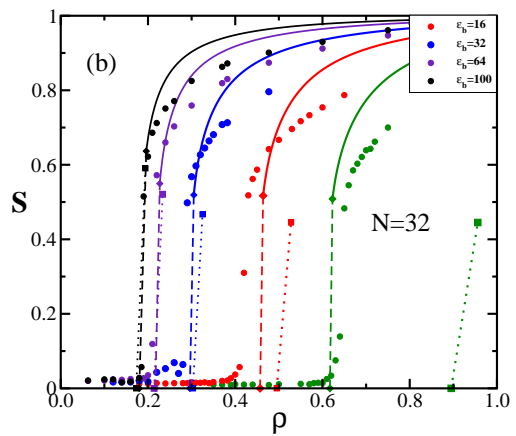}

\caption{Nematic order parameter $S$ from MD (filled circles) vs
concentration $\rho$ for $N=32$ and various choices of $\epsilon_b$, as
indicated. Full curves denote corresponding predictions of DFT-CS.~\cite{notedft}  
$I-N$ coexistence is indicated by diamonds and broken straight lines (lever rule). 
Corresponding predictions from Chen~\cite{chen93} are shown by squares and dotted
lines.} 
\label{fig3}
\end{figure}

However, most interesting is the qualitative discrepancy between MD and DFT with respect to the density dependence of the order parameter $S$ (Fig.~\ref{fig3}), whereby the DFT result approaches saturation much faster than according to MD, that is, DFT
significantly overestimates the degree of ordering in the nematic phase. We
attribute this fact to the neglect of long wavelength fluctuations in the
nematic phase, reflecting the mean-field character of the DFT. The situation is
analogous to the case of the molecular field approximation (MFA) for an
isotropic Heisenberg ferromagnet: the  MFA also does not allow for effects due
to magnons. Both in this case and at the $I-N$ transition, a continuous
symmetry is broken, but for semiflexible polymers the situation is special since
an additional lengthscale (the deflection length) matters.

\begin{figure}[htb]
\includegraphics[scale=0.3]{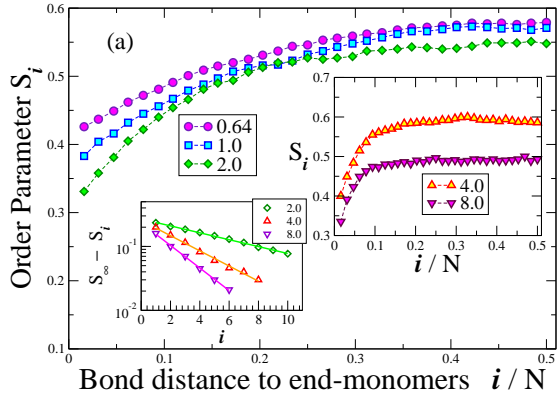}

\vspace*{0.90cm}
\includegraphics[scale=0.3]{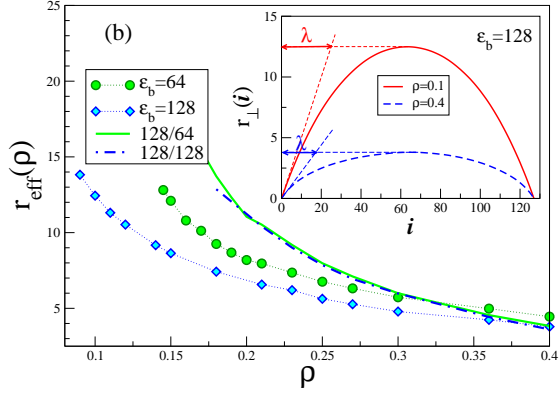}
\caption{(a) Local order parameter $S_i$ (referring to bond vector $\vec{a}_i$)
plotted vs $i/N$ and averaged over all equivalent bonds in the system for the
case $N=64$ and several choices of $N/\epsilon_b$, as indicated. The left inset
shows a semi-log plot of $S_\infty - S_i$ vs $i$ so as to demonstrate
Eq.~\ref{eq4}. From the slope the deflection length $\lambda$ is extracted as
$\lambda / \ell_p = 2.36,\; 3.65$, and $8.2$ for $\epsilon_b = 8,\; 16$, and
$32$, respectively. (b) Variation of the confinement radius $r_{eff}$ with
concentration $\rho$ for semiflexible chains with $N=128$ and two degrees of
stiffness $\epsilon_b = 64$ and $128$, computed from Eq.~(\ref{eq4}) and from
the maximum of $\langle (r_{i,\bot}-r_{j,\bot})^2\rangle$.  The inset shows the
mean-squared displacement of consecutive beads, $\langle (r_{i,\bot} -
r_{j,\bot})^2 \rangle$, perpendicular to the respective end-to-end vector
$\vec{R}_e$, averaged over all chains in the nematic phase for $\epsilon_b=128$
and two densities.}
\label{fig4}
\end{figure}

{\em Deflection length and cylindrical confinement}.---Unlike nematic order of rigid rods, the local order parameter $S_i$ along the
contour of an individual chain is non-uniform (Fig.~\ref{fig4}a), and can be
described by
\begin{equation}
S_{\infty}-S(i)\propto \exp(-i\ell_b/\lambda),
\label{eq3}
\end{equation}
$S_{\infty}$ being the order parameter in the center of a chain (for $L
\rightarrow \infty$), and $\lambda$ can be taken as a definition of the
deflection length~\cite{odijk85,odijk86,chen93}. Alternatively, we can measure
the mean-square monomer displacement $\langle (\vec{r}_{i,\bot} -
\vec{r}_{j,\bot})^2 \rangle$ in the direction perpendicular to the end-to-end
vector $\vec{r}_N - \vec{r}_1$ as function of the bead index (Fig.~\ref{fig4}b,
inset). In the nematic phase this displacement increases linearly with $s=j-i$ and 
reaches a flat maximum 
(of height $r_{eff}^{2}$) at distance $\lambda$ along the contour. The deflection length is
normally~\cite{odijk83,yang07,chen13,chen15} defined for a semiflexible polymer
confined in a cylinder of radius $r_{eff}$.

Considering the initial growth of the mean-squared angle with the distance $s$
along the contour, $\langle\theta^2(s)\rangle=2s\ell_b/\ell_p$, and equating
this to $r_{eff}^{2}/\lambda^2$ for $s\ell_b=\lambda$, one concludes that
$\lambda=(\ell_pr_{eff}^{2})^{1/3}$ and $1-S\approx 3/2\langle\theta^2\rangle
\approx 3/2\left(r_{eff}/\ell_p\right)^{2/3}$. Since the average projection of
each bond along the cylinder axis is
$\ell_b\langle\cos\theta\rangle\approx\ell_b(1-\langle\theta^{2}\rangle/2)$, the
reduction of the mean-squared end-to-end distance becomes $1-\langle
R_{e}^{2}\rangle^{1/2}/L\approx\langle\theta^2\rangle/2
=\left(\frac{r_{eff}}{\ell_p}\right)^{2/3}/2$. These scaling arguments can be
made more precise to yield,~\cite{yang07,chen13,chen15} for $L/\ell_{p}\gg 1$,

\begin{equation} 
1-S=3\left(1-\frac{\sqrt{\langle
R_{e}^{2}\rangle}}{L}\right)=0.51\left(\frac{2r_{eff}}{\ell_p}\right)^{2/3}
=3\frac{\lambda}{\ell_p}.
\label{eq4}
\end{equation}

\begin{figure}[htb]
\includegraphics[scale=0.3]{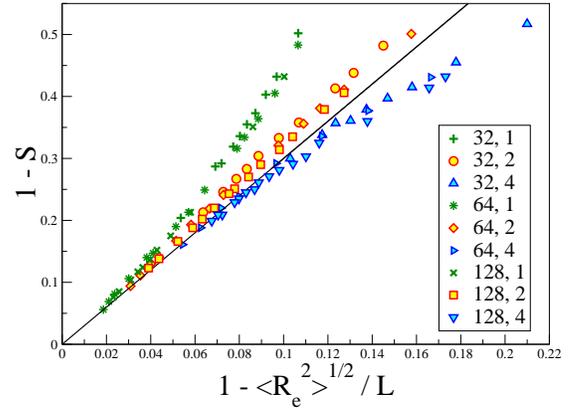}

\caption{Plot of $1 - S$ vs the relative reduction $1 - \langle R_e^2
\rangle^{1/2}/L$ of the end-to-end distance for three choices of $N = 32,\; 64,$
and $128$, and three choices of the $N/\epsilon_b = 1,\; 2$, and $4$, as
indicated. Different points with the same symbol refer to different choices of
the density $\rho$. The fully stretched chain would be the origin of the plot
whereas the straight line shows Eq.~(\ref{eq4}). Rigid rods would correspond to
the ordinate axis here.}
\label{fig5}
\end{figure}

We now suggest that nematic order of semiflexible polymers can be essentially
understood in terms of cylindrical confinement as a collective effect of the
neighboring chains of the considered chain (Fig.~\ref{fig1}c). These cylinders
in Fig.~\ref{fig1}c must not be confused with the tubes due to entanglements in
solutions of semiflexible polymers, controlling the viscoelastic dynamics in the
isotropic phase~\cite{morse98,ramanathan07,hinsch07,glaser11}. 
Fig.~\ref{fig4}b shows how both
$\lambda$ and $r_{eff}$ can be extracted from the data. Choosing different
values of $N$ and $\epsilon_b$, we can also test the left part of
Eq.~(\ref{eq4}), see Fig.~\ref{fig5}. For the regime where Eq.~(\ref{eq4})
should hold, namely $L/\ell_{p}\gg 1$ and $\langle\theta^2\rangle\ll 1$, i.e.
$1-S\le 0.2$, we get very good agreement with no adjustable parameters
whatsoever. For $L/\ell_{p}\le 1$, the data display curvature and bend upwards
away from a straight line. This is expected, of course, since for $L/\ell_{p}<
1$ the end-to-end distance of such rather stiff ``flexible rods'' cannot
decrease much. Disordering of the nematic phase then occurs predominantly due to
misorientation of the flexible rods relative to the director as a whole.
Remarkably, different choices of $L$ and $\ell_p$ in the representation of
Fig.~\ref{fig5} yield a set of master curves depending on the single parameter
$L/\ell_p$ only. However, different scaled concentrations $\rho \ell_p/d$ for a
given $L/\ell_p$ do not coincide on the same point of the master curve, but
differ systematically. Using the result $1-S=3(\lambda/\ell_p)$, we obtain
alternative estimates for $\lambda$. For the cases shown in Fig.~\ref{fig3}a, we
thus find for $\epsilon_b=8,\; 16,$ and $32$ the values $\lambda = 1.36,\;
2.14,$ and $4.7$, respectively. These estimates are systematically somewhat
smaller than those extracted from Fig.~\ref{fig4}a via Eq.~(\ref{eq3}), but
exhibit a similar trend.

This analysis in terms of cylindrical confinement does not mean that the nearest
neighbors of a chain enclose it in a cylinder of radius $r_{eff}$
(Fig.~\ref{fig1}c), rather this cylinder is shared by many chains. This is
readily seen when we compute a radius of a cylinder from $\rho$ via $r_\rho =
[N/(\pi\rho\langle R_{e}^{2} \rangle^{1/2}) ]^{1/2}$, i.e., a cylinder
containing the monomers of one chain only (and solvent particles). For
concentrated solutions $r_{\rho}$ is comparable to $\sigma$, of course, see
Fig.~\ref{fig4}b, while $r_{eff}$ extracted from Eq.~(\ref{eq4}) is much larger
(it increases proportional to $\ell_p$). This description implies
Fig.~\ref{fig1}c, i.e., each chain is confined in a tube of radius $r_{\rho}$,
but this tube as a whole is like a wormlike chain, making excursions of order of
$r_{eff}$ on a length scale $\lambda$ along the cylinder axis. Since the
cylinder of  radius $r_{eff}$ contains a bundle of chains (which may be twisted
around each other, a feature missed in our two-dimensional cartoon), it is clear
that the deflections of these chains sharing one cylinder are coherent
collective excitations, because the tubes of radius $r_{\rho}$ must be
essentially space-filling. Of course, for a semidilute solution $r_{\rho}$ may
exceed $\sigma$ considerably, and then the chains have additional bending
degrees of motion within their individual tubes as indicated qualitatively in
Fig.~\ref{fig1}c.

{\em Conclusions}.---In summary, we have shown that the nematic phase of
semiflexible polymers exhibits {\em collective deflection modes} on length scale
$\lambda$ of amplitude $r_{eff}$ perpendicular to the director, if $L/\ell_{p}
\gg 1$, and both $\lambda$ and $r_{eff}$ can be directly predicted from the
order parameter $S$ (Eq.~(\ref{eq4})). We feel that the picture of nematic order
of semiflexible polymers developed here has also important implications for both
linear and nonlinear elastic response of such systems, and corresponding
experiments testing our ideas would be very welcome. So far the deflection
length has only been measured for a semiflexible chain in a nematic
solvent~\cite{dogic04}. There one might need to consider defects such as
hairpins in the structure (which have been occasionally detected in our
simulations). Note that in nematics formed from rigid rods each rod is confined
in a cylinder of radius $r_\rho$ while the length scales $\lambda$ and $r_{eff}$
do not exist!  For the interpretation of the experiments, the version of DFT
that we have used and validated here could be very useful, since it can be
worked out for a much wider parameter range compared to MD. Our study clearly
shows limitations of the previous theories of the $I-N$ transition of
semiflexible polymers, see, e.g.,~Fig.~\ref{fig2}b, and should provide a better
understanding of experiments. It would be very interesting to study the
corresponding static and dynamic collective structure factors, but this is
beyond our scope here. It remains a challenge to extend the analytic
theories~\cite{khokhlov81,khokhlov82,odijk85,odijk86,chen93,chen13} to
self-consistently predict the length scales $\lambda$ and $r_{eff}$ from the
molecular parameters $\ell_p$, $d$ and $L$ and the polymer density.

{\em Acknowledgements}

S.A.E. acknowledges financial support from the Alexander von Humboldt
Foundation. A.M. thanks for partial support under the grant No $BI314/24$. We
are particularly indebted to Dr. P. Virnau for his help and advice with GPU
computing.
Parts of this research were conducted using the supercomputer Mogon and/or advisory services offered by Johannes Gutenberg University Mainz (www.hpc.uni-mainz.de), which is a member of the AHRP and the Gauss Alliance e.V.
The authors gratefully acknowledge the computing time granted on the supercomputer Mogon at Johannes Gutenberg University Mainz (www.hpc.uni-mainz.de).


\end{document}